\documentclass[useAMS,usenatbib]{mn2e}

\newif\ifAMStwofonts
\usepackage{epsfig}

\title[C$_{60}$/anthracene adducts]{Infrared spectroscopy of fullerene C$_{60}$/anthracene
 adducts}
\author[D. A.~Garc\'{\i}a-Hern\'andez et al.]{D. A.~Garc\'{\i}a-Hern\'andez$^{1,2}$, F. Cataldo$^{3,4}$ and A. Manchado$^{1,2,5}$\\
$^{1}$Instituto de Astrof\'{\i}sica de Canarias, V\'{\i}a L\'actea s/n, E-38200 La Laguna, Spain; agarcia@iac.es\\
$^{2}$Departamento de Astrof\'{\i}sica, Universidad de La Laguna (ULL), E-38206 La Laguna, Spain\\
$^{3}$INAF- Osservatorio Astrofisico di Catania, Via S. Sofia 78, Catania 95123, Italy\\
$^{4}$Actinium Chemical Research srl, Via Casilina 1626/A, 00133 Rome, Italy\\
$^{5}$Consejo Superior de Investigaciones Cient\'{\i}ficas, Madrid, Spain}

\begin{document}

\date{Accepted 2013 xx xx. Received 2012 xx xx; in original form 2012 xx xx}

\pagerange{\pageref{page}--\pageref{lastpage}} \pubyear{2013}

\maketitle

\label{firstpage}

\begin{abstract} 
Recent {\it Spitzer Space Telescope} observations of several astrophysical
environments such as Planetary Nebulae, Reflection Nebulae, and R Coronae
Borealis stars show the simultaneous presence of mid-infrared features
attributed to neutral fullerene molecules (i.e., C$_{60}$) and polycyclic
aromatic hydrocarbons (PAHs). If C$_{60}$ fullerenes and PAHs coexist in
fullerene-rich space environments, then C$_{60}$ may easily form adducts with a
number of different PAH molecules; at least with catacondensed PAHs. Here
we present the laboratory infrared spectra ($\sim$2$-$25 $\mu$m) of C$_{60}$
fullerene and anthracene Dies-Alder mono- and bis-adducts as produced by
sonochemical synthesis. We find that C$_{60}$/anthracene Diels-Alder adducts
display spectral features strikingly similar to those from C$_{60}$ (and
C$_{70}$) fullerenes and other unidentified infrared emission features. Thus,
fullerene-adducts - if formed under astrophysical conditions and
stable/abundant enough - may contribute to the infrared emission features
observed in fullerene-containing circumstellar/interstellar environments. 
\end{abstract}

\begin{keywords}
astrochemistry; circumstellar matter; ISM: molecules; methods: laboratory; techniques: spectroscopic
\end{keywords}

\section{Introduction}

\begin{figure*}
\epsfxsize=18truecm
\epsffile{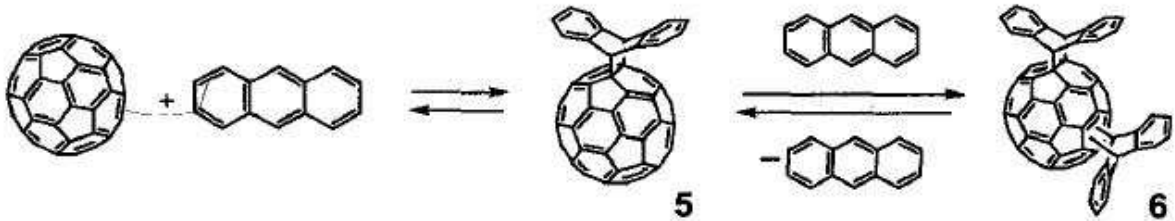}
\caption{Scheme of the addition reaction of anthracene to C$_{60}$
fullerene. It is schematized the fact that initially anthracene forms a
mono-adduct (\#5) but in an excess of anthracene, a bis-adduct (\#6) is obtained
(Komatsu et al. 1999).}
\end{figure*}

Fullerene molecules (e.g., C$_{60}$ and C$_{70}$) were first synthesized at
laboratory in an effort to unveil the formation mechanism of long-chain carbon
molecules in interstellar and circumstellar media (Kroto et al. 1985). The
possible existence of fullerenes in astrophysical environments has been a
controversial issue until recently when these complex organic molecules were
identified in a young Planetary Nebula (Tc 1; Cami et al. 2010). Based on the
lack of the classical aromatic infrared bands (AIBs) (e.g., at 3.3, 6.2, 7.7,
8.6, and 11.3 $\mu$m) usually attributed to polycyclic aromatic hydrocarbons
(PAHs; e.g., Leger \& Puget 1984), Cami et al. (2010) proposed that fullerenes
flourish in the H-deficient and C-rich inner regions of Tc 1. This intepretation
was in agreement with the original laboratory studies on the formation of
fullerenes (Kroto et al. 1985; de Vries et al. 1993) that show that fullerenes
at laboratory are efficiently produced in the absence of hydrogen. However,
Garc\'{\i}a-Hern\'andez et al. (2010) reported the simultaneous detection of
C$_{60}$ fullerenes and PAHs in several Planetary Nebulae (PNe) with normal H
abundances. In addition, the unexpected detection of C$_{60}$ fullerenes
together with PAHs only in the two least H-deficient R Coronae Borealis stars DY
Cen and V854 Cen (Garc\'{\i}a-Hern\'andez et al. 2011a,b) confirmed the
surprising results obtained in PNe. This challenged our understanding of the
fullerenes formation in space, showing that, contrary to general expectation,
fullerenes are efficiently formed in H-rich circumstellar environments only.
Indeed, the C$_{60}$ fullerene has since then been detected also in a
variety of environments such as the interstellar medium (Sellgren et al. 2010),
a post-asymptotic giant branch (post-AGB) star and post-AGB circumbinary discs
(Zhang \& Kwok 2011; Gielen et al. 2011), the Orion nebula (Rubin et al. 2011),
a Herbig Ae/Be star and young stellar objects (YSOs; Roberts et al. 2012).

The detection of fullerenes around evolved stars such as our Sun indicates that
these complex molecules are much more common and abundant in
circumstellar/interstellar environments than was originally believed. It
reinforces the idea that fullerenes and molecular-related species are ubiquitous
in the interstellar medium, playing an important role in many aspects of
circumstellar/interstellar Chemistry. In addition, fullerenes and PAHs may be
mixed in the circumstellar envelopes of (proto-) PNe (e.g.,
Garc\'{\i}a-Hern\'andez et al. 2010, 2011c, 2012a; Zhang \& Kwok 2011), raising
the exciting possibility of forming adducts of fullerenes with PAHs.

C$_{60}$ fullerene is an electron deficient polyolefin, which is able to give
adducts with a number of different molecules (e.g., Yurovskaya \& Trushkov 2002;
Hirsch \& Brettreich 2005). In particular, C$_{60}$ fullerene can react with
catacondensed PAHs (e.g., acenes such as anthracene, pentacene) to form
fullerene/PAH adducts via Dies-Alder cycloaddition reactions (Briggs \& Miller
2006). It is to be noted here that the Diels-Alder reaction between
fullerenes and pericondensed or very large PAHs has not been explored
experimentally and the possible formation of such more complex fullerene/PAH
adducts is still an open issue. The formation of a Diels-Alder adduct of
C$_{60}$ with anthracene is the simplest fullerene/PAH adduct and was one of the
first [4+2] addition reactions discovered where the former molecule is acting as
a dienophile and the latter as a diene (e.g., Komatsu et al. 1993a,b). C$_{60}$
fullerene can react with anthracene and form the C$_{60}$/anthracene mono-adduct
and the bis-adduct. Figure 1 displays an addition reaction scheme of the
C$_{60}$ fullerene with anthracene. C$_{60}$ and anthracene react under
different reaction condition to form the C$_{60}$/anthracene mono-adduct (adduct
\#5 in Fig.1) and, in case of anthracene excess also the trans$-$1 bis$-$adduct
(adduct \#6 in Fig. 1). Different yields of the C$_{60}$/anthracene mono- and
bis-adducts can be obtained at laboratory depending on the method employed in
the production process (see e.g., Cataldo et al. 2013 for a review). The
efficiency of the cycloaddition of anthracenes and other acenes to C$_{60}$ can
be improved by performing the reaction under the conditions of high speed
vibration grinding of solid reagents, known as mechanochemical synthesis (e.g.,
Komatsu et al. 1999; Wang et al. 2005). Inspired by this mechanochemical
synthesis we have recently been stimulated to explore the sonochemical synthesis
of these C$_{60}$/anthracene adducts (Cataldo et al. 2013). 

The interest in the C$_{60}$/anthracene and other fullerene/PAHs adducts regards
the access to new fullerene-related molecules, which may be present in the
circumstellar and interstellar environments where C$_{60}$ fullerenes have
already been found (see above). For example, the posible presence of other
complex fullerene-based molecules (e.g., fullerenes bigger than C$_{60}$ or
multishell fullerenes) in fullerene-containing space environments is
strengthened by a recent study of the diffuse interstellar bands (DIBs) in PNe
with fullerenes (Garc\'{\i}a-Hern\'andez \& D\'{\i}az-Luis 2013). In addition,
C$_{60}$ fullerenes seem to be mixed with PAHs and the present debate involves
not only how and where fullerene was formed but also if C$_{60}$ may interact
with PAHs forming adducts. In this paper we show that  C$_{60}$ is not
inert toward PAHs and under certain circumstances (e.g., where the two reagents
co-exist) they can react forming new products where fullerene cage is
functionalized by PAHs moieties attached on it by covalent bonds. We report the
laboratory infrared spectra ($\sim$2$-$25 $\mu$m) of C$_{60}$/anthracene
Dies-Alder mono- and bis-adducts - the simpler example of a family of
fullerene/PAHs adducts - as produced by sonochemical synthesis. These
laboratory spectra are compared with the {\it Spitzer Space Telescope} spectra
of fullerene PNe in an attempt to identify some still unidentified mid-IR
features displayed by some of these sources.

\section{Sonochemical synthesis of C$_{60}$/anthracene adducts}

C$_{60}$ fullerene was 99\% pure grade and was obtained from MTR Ltd (USA).
Anthracene and all the solvents used were obtained from Sigma-Aldrich (Germany
and USA). The Fourier-Transform infrared (FT-IR) spectra (at a spectral
resolution of $\sim$600) were recorded on a IR300 spectrometer from
Thermo-Fischer in transmittance mode with samples embedded in KBr. The
infrared band shift and band broadening effects by passing from KBr matrix to
solid Ar matrix and to the gas phase spectra are relatively small (see e.g.,
Iglesias-Groth et al. 2011 for the case of the C$_{60}$ and C$_{70}$
fullerenes). Consequently, also for the C$_{60}$/anthracene adducts we do not
expect important differences from our spectra measured in KBr in the lab and
those which could derive from the adduct in different environmental conditions
than those used in the lab.

The ultrasonic bath used for the synthesis was a Branson 1510. The reaction
between C$_{60}$ and anthracene (at 1:1 and 1:2 initial molar ratio) has been
successfully conducted by us under sonication in a ultrasonic bath using a
benzene solution (see Cataldo et al. 2013 for more details). In short, C$_{60}$
and anthracene  were dissolved in a conical flask at room temperature and the
flask was suspended in the ultrasonic bath filled with water. The sonication was
run for 5 h and the grey reaction product was recovered by distillation of
benzene under reduced pressure in a water bath kept at 338 K. Our detailed
thermogravimetric analysis (TGA) and differential thermal analysis (DTA)
(Cataldo et al. 2013) show that the grey C$_{60}$/anthracene reaction mixture is
composed by 33\% of the mono-adduct and by 66\% of the trans-1 bis-adduct that
are shown in the addition reaction scheme of Figure 1 (see Section 3.1). 
The synthesis of the C$_{60}$/anthracene adducts is made starting from pure
C$_{60}$ and pure anthracene and the undesired by-products in the adduct
reaction were removed during the purification process (Section 3.2; see also
Cataldo et al. 2013 for more details). Thus, there are no other contaminants
available even as minor impurities.

From literature data it is known that the mechanochemical synthesis of
C$_{60}$/anthracene adducts in the solid state can reach a 55\% yield in 1 h
(e.g., Komatu et al. 1999; Wang et al. 2005) with the mono-adducts and
bis-adducts representing 4/5 and 1/5 of the mixture, respectively\footnote{Note
that these previous works on C$_{60}$/anthracene adducts permit us to clearly
identify our products.}. However, the main product of our novel sonochemical
synthesis of C$_{60}$/anthracene adducts is the trans-1 bis-adduct (Cataldo et
al. 2013). In addition, such a product can be obtained almost pure after
appropriate purification of the reaction mixture (see below). Thus, the
sonochemical synthesis gives a direct access to pure C$_{60}$/anthracene trans-1
bis-adducts (see Sections 3.2 and 4). 

The grey 1:1 and 1:2 adducts (51 mg and 52 mg, respectively) were exhaustively
extracted in a Soxhlet apparatus for 6h using n-hexane as extracting solvent.
This causes the extraction of the mono-adduct from the mixture and at the end of
the extraction process, the n-hexane solutions contained free unreacted C$_{60}$
and unreacted anthracene (Cataldo et al. 2013). The bis-adducts are instead
insoluble in n-hexane and were recovered in 26.5 mg and 26.8 mg after drying for
the 1:1 and 1:2 initial molar ratio, respectively. Thus, the yields of both
adducts over the grey reaction product was $\sim$50\%.

\begin{figure}
\epsfxsize=8truecm
\epsffile{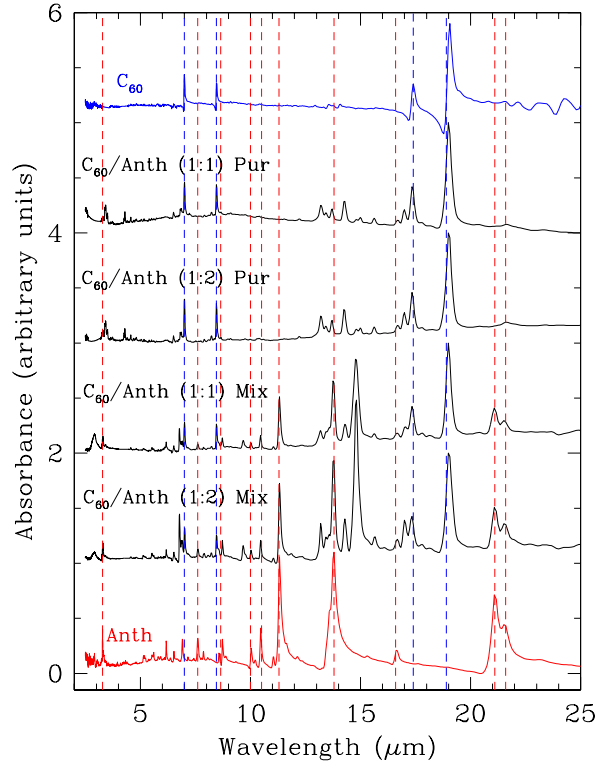}
\caption{FT$-$IR normalised spectra in KBr. From top to bottom: C$_{60}$
reference spectrum; purified C$_{60}$/anthracene (1:1 initial molar ratio);
purified C$_{60}$/anthracene (1:2 initial molar ratio); grey reaction mixture of
C$_{60}$/anthracene (1:1 initial molar ratio); grey reaction mixture of
C$_{60}$/anthracene (1:2 initial molar ratio); anthracene reference spectrum.
Note that the neutral C$_{60}$ (dotted) and Anthracene (dashed) band positions
are marked.} 
\end{figure}

\section{Laboratory spectra of C$_{60}$/anthracene adducts}

\subsection{Grey reaction product of C$_{60}$/anthracene adducts}

The FT-IR spectra of the grey, unpurified reaction mixtures recovered
(irrespective of the initial molar ratio of the reactants adopted) show a series
of infrared  bands, which are those of free C$_{60}$ and free anthracene (see
Figure 2). In Figure 2 it is possible to distinguish the characteristic four
band pattern of pure C$_{60}$ at $\sim$7.0, 8.5, 17.4, and 18.9 $\mu$m (or at
1426, 1180, 575, and 525 cm$^{-1}$). Furthermore, the strongest infrared bands
due to free anthracene can be clearly distinguished as well (see Figure 2); for
example those at 11.3, 13.8, 21.1, and 21.6 $\mu$m (or at 882, 725, 475, and 464
cm$^{-1}$). 

The rest of spectral features seen in the FT-IR spectra of the
C$_{60}$/anthracene reaction mixtures (composed by 33\% of the mono-adduct and
by 66\% of the trans-1 bis-adduct) are identical to those displayed by the
purified trans-1 bis-adducts and that are discussed in the next Section.
However, it is worth mentioning that a few infrared bands (e.g., at 14.8 $\mu$m)
in the unpurified reaction mixtures appear stronger than in the purified
C$_{60}$/anthracene bis-adducts (Figure 2). This is likely due to the
extra-contribution from the C$_{60}$/anthracene mono-adduct at these 
wavelengths.

\subsection{Purified C$_{60}$/anthracene bis-adducts}

\begin{figure}
\epsfxsize=8truecm
\epsffile{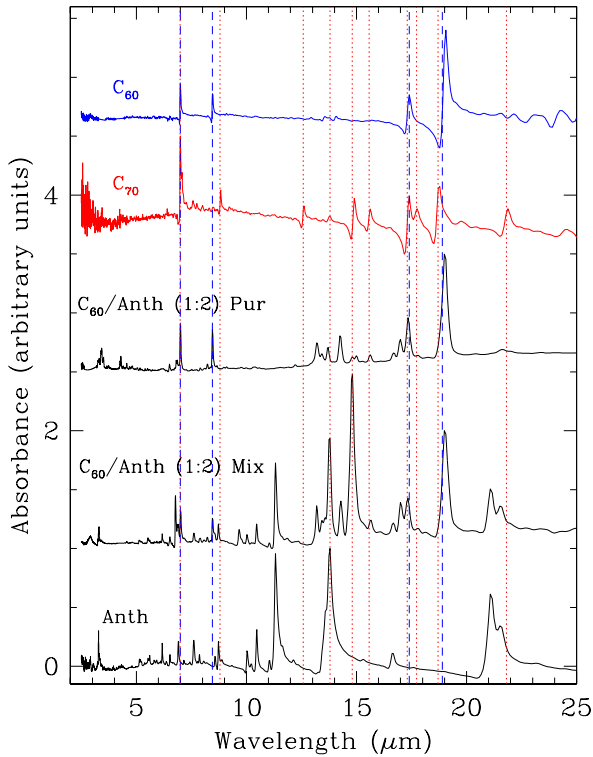}
\caption{FT$-$IR normalised spectra in KBr. From top to bottom: C$_{60}$
reference spectrum; C$_{70}$ reference spectrum; purified C$_{60}$/anthracene 
(1:2 initial molar ratio); grey reaction mixture of C$_{60}$/anthracene (1:2
initial molar ratio); anthracene reference spectrum. Note that the neutral
C$_{60}$ (dotted) and C$_{70}$ (dashed) band positions are marked.} 
\end{figure}

As we have mentioned before, the purification of the grey
C$_{60}$/anthracene mixture with n-hexane causes the extraction of the
mono-adduct from the mixture (together with the extraction of the not reacted
C$_{60}$ and anthracene). The residue of the extraction is essentially
constituted by almost pure trans-1 bis-adducts (Figure 1, adduct \#6). 

Surprisingly, the purified C$_{60}$/anthracene bis-adducts obtained through the
sonication of the 1:1 and 1:2 initial molar ratio mixtures display an identical 
FT-IR spectrum (Figure 2). This strongly suggests that both C$_{60}$/anthracene
bis-adducts have the same chemical structure. Clear evidences of the adduct
formation derive from the C-H infrared bands displayed by the adduct and due to
the cycloalyphatic nature of the C-H bonds connecting the anthracene molecule
with the fullerene cage (see Figure 1). In the adduct C$_{60}$ maintains its
individuality and its cage structure while anthracene being a dienophile loses
the aromaticity of the central ring of the molecule forming a cycloaliphatic
structure. However, the other two anthracene rings are still aromatic and
display the typical aromatic bands in the infrared spectrum. It can be observed
that there are a total of 4 aliphatic C-H groups equivalent in couple. Indeed,
the FT-IR spectra of the purified bis-adducts (see Figure 2) show a series of
bands at 3.39, 3.43, and 3.52 $\mu$m (or at 2842, 2917, and 2947 cm$^{-1}$),
which we tentatively attribute to aliphatic C-H stretching vibrations (see
below) and that are not present in the neutral C$_{60}$ spectrum nor in the
anthracene spectrum (Figure 2). Indeed, the anthracene molecule displays the
aromatic C-H stretching at 3.28 $\mu$m (3051 cm$^{-1}$) only. The
out-of-plane bending vibration at about 13.8 $\mu$m of anthracene is still
observable in the bis-adducts (due to the presence of four aromatic C-H groups
at one ring), although this band is much more intense in the grey reaction
products spectrum. Also, the 11.3 $\mu$m band (due to isolated H bound to an
aromatic ring) is not present in the pure bis-adducts infrared spectrum, as
expected. However, some spectral features seen in our laboratory spectra of the
purified C$_{60}$/anthracene bis-adducts may not have a straightforward
interpretation. In the adduct we have only two tertiary C-H bands in addition to
aromatic C-H. From two identical tertiary C-H bonds, we would expect the
appearance of one up to two C-H stretching bands shifted to longer wavelengths
compared to the bands of CH$_{2}$ and CH$_{3}$ groups (see Allamandola et al.
1992), something that it seems to be not observed. In addition, the weak and
broad 3.3 $\mu$m band is difficult to understand since we still have eight
aromatic C-H groups in the adduct (compared to only two aliphatic C-H). A
theoretical calculation of the C$_{60}$/anthracene IR spectrum (although out of
the scope of this paper) would be very useful to completely understand the
laboratory spectra and band assignments for a structure like that of
C$_{60}$/anthracene adducts. 

Interestingly, the purified C$_{60}$/anthracene bis-adducts show also strong
spectral features which are coincident with those of neutral C$_{60}$ (e.g., at
$\sim$7.0, 8.4, 17.3, and 19.0 $\mu$m; Iglesias-Groth et al. 2011) - even the
relative strenghts are strikingly similar (Figure 2). The presence of the
$\sim$7.0, 8.4, 17.3, and 19.0 $\mu$m C$_{60}$ bands in our C$_{60}$/anthracene
bis-adducts FT-IR spectrum gives additional evidence that the fullerene
cage is completely intact in the adduct while there are no more bands
attributable exclusively to free anthracene as instead happened in the case of
the unpurified product. 

Moreover, the purified C$_{60}$/anthracene bis-adducts also show a series of new
infrared absorption bands, which are not present in the spectra of pure C$_{60}$
or pure anthracene. The latter IR bands are located at $\sim$6.5, 6.8/6.9, 13.2,
13.5, 13.8\footnote{Note that the 13.8 $\mu$m band may be due to the 
out-of-plane bending vibration of anthracene (see above).}, 14.3, 14.9, 15.1,
15.6, 17.0, 17.8, and 21.6 $\mu$m. Indeed, some of these new IR bands seem to be
coincident with the known transitions of the neutral C$_{70}$ fullerene (e.g.,
at $\sim$7.0, 13.8, 14.9, 15.6, 17.3, 17.8, and 18.9 $\mu$m; Iglesias-Groth et
al. 2011). This is shown in Figure 3, where we display the FT-IR spectrum of the
purified C$_{60}$/anthracene bis-adducts (1:2 initial molar ratio) together with
the neutral C$_{70}$ reference spectrum. As we have mentioned before, our
synthesis and purification process gives almost pure trans-1 bis-adducts, and
therefore also C$_{70}$ is not available even as a minor impurity. 

\section{Formation of fullerenes and fullerene/PAHs adducts}

The formation route of fullerenes in evolved stars such as PNe and in the
interstellar medium is still not fully understood. Several scenarios for
fullerene formation have been proposed, including: i) photochemical processing
of hydrogenated amorphous carbon grains (HACs; Garc\'{\i}a-Hern\'andez et
al. 2010);  ii) destruction of large PAHs by shocks (Cami et al. 2011); iii)
photochemical processing of PAHs where PAHs are converted into graphene, and
subsequentely fullerenes (Bern\'e \& Tielens 2012)\footnote{Bettens \&
Herbst (1996) proposed and alternative route in the interstellar medium through
ion/molecule synthesis in which previously formed tricyclic rings can be
converted to fullerenes by collisional processes.}  

The coexistence of fullerenes and PAHs - as well as of other molecular species
such as HACs, PAH clusters, and small dehydrogenated carbon clusters (planar
C$_{24}$ or proto-graphene) - in the infrared spectra of PNe with
fullerenes (Garc\'{\i}a-Hern\'andez et al. 2011c, 2012a) strongly supports the
laboratory experiments carried out by Scott and colleagues in the nineties,
which showed that the decomposition of HACs is sequential with small
dehydrogenated PAH molecules being released first, followed by fullerenes and
large PAH clusters (Scott et al. 1997). The most recent studies about
fullerenes in PNe (Bernard-Salas et al. 2012; Garc\'{\i}a-Hern\'andez et al.
2012a; Micelotta et al. 2012) seem to support the HAC's photochemical processing
(e.g., by UV irradiation) originally proposed by Garc\'{\i}a-Hern\'andez et al.
(2010) as the most likely fullerene formation route. In particular, Micelotta et
al. (2012) present a good review on the possible fullerene formation processes
in space, showing for example that the top-down fullerene formation scenario by
Bern\'e \& Tielens (2012) cannot probably work in PNe and concluding that
formation of fullerenes in PNe likely starts from HAC processing.

It is worth mentioning that the spatial distribution of the 16.4 $\mu$m
emission (and other PAH-like features in the 15-20 $\mu$m range) is found to be
different to that of the 18.9$\mu$m C$_{60}$ emission in the reflection nebula
NGC 7023 (Sellgren et al. 2010; Bern\'e \& Tielens 2012; Peeters et al. 2012).
This fact has been used to favor the destruction of large PAHs for fullerene
formation, either by shocks (Cami et al. 2011)\footnote{Note that the same
authors favor the HAC's scenario in their subsequent papers Bernard-Salas et al.
(2012) and Micelotta et al. (2012).} or by UV irradiation (Bern\'e \& Tielens
2012) against the HAC's scenario. As far as we know there is no firm
identification of the carrier of the 16.4 $\mu$m feature (the same holds for the
other features in the 15-20 $\mu$m range) although the latter authors attribute
this feature to PAHs. For example, Duley \& Hu (2012) show that HAC
nano-particles also display a feature at 16.4 $\mu$m and they attribute this
feature to proto-fullerenes. In addition, the 16.4 $\mu$m feature is lacking in
all PNe where fullerenes have been detected so far (Garc\'{\i}a-Hern\'andez et
al. 2012a). Thus, one can not extrapolate the results in NGC 7023 to PNe.
Furthermore, PAHs are seen everywhere in NGC 7023 (see Figure 1 in Berne \&
Tielens 2012); the only difference is that fullerenes and PAHs appear to coexist
in the same location close to the central star, while only PAHs are detected
further away from the star (Berne \& Tielens 2012; Micelotta et al. 2012).
Actually, the unique information on the spatial distribution of fullerenes and
other dust species in PNe is coming from Bernard-Salas et al. (2012). The latter
authors presented tentative evidence (at a marginal spatial resolution of
2"/pixel) that the 8.5 $\mu$m emission (and attributed to C$_{60}$) in Tc 1 is
extended and peaks at 2-3 pixels from the central star while the 11.2 $\mu$m
emission peaks at a similar distance to the other side of the star. It should be
noted here that both emissions coexist in the nebula but they peak at different
places. Indeed, fullerenes could be formed in clumps and the tentative evidence
by Bernard-Salas et al. (2012) may point to this. Mid-IR images at much higher
spatial resolution (e.g., at sub-arcsecond level) are desirable before reaching
a conclusion about the relative spatial distribution of fullerenes and PAHs in
Tc 1 and other PNe.

\begin{figure}
\epsfxsize=8truecm
\epsffile{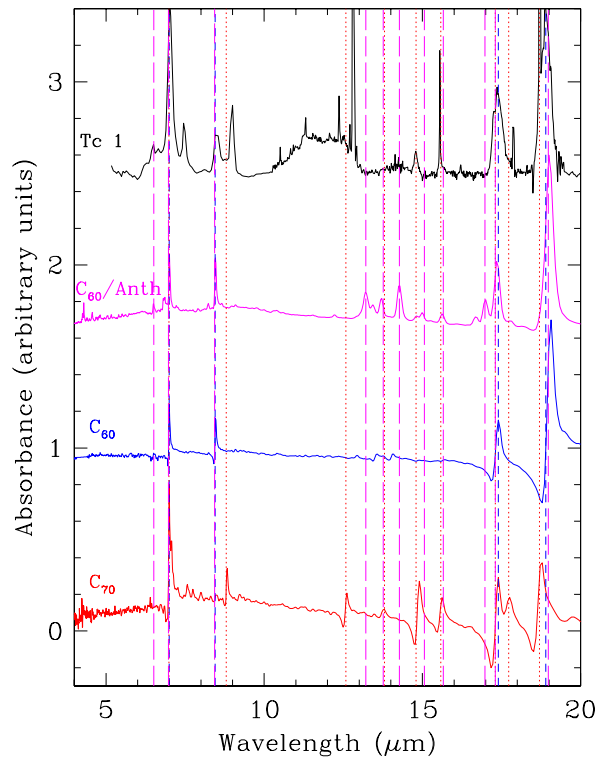}
\caption{Dust continuum subtracted spectra ($\sim$5$-$20 $\mu$m) of PN Tc 1
(Garc\'{\i}a-Hern\'andez et al. 2010) in comparison with the  purified
C$_{60}$/anthracene adducts (1:2 initial molar ratio) FT-IR spectrum (magenta)
and the reference neutral C$_{60}$ (blue) and C$_{70}$ (red) spectra (see e.g.,
Iglesias-Groth et al. 2011). Note that the laboratory spectra are normalised and
displaced for clarity.} 
\end{figure}

\begin{figure}
\epsfxsize=8truecm
\epsffile{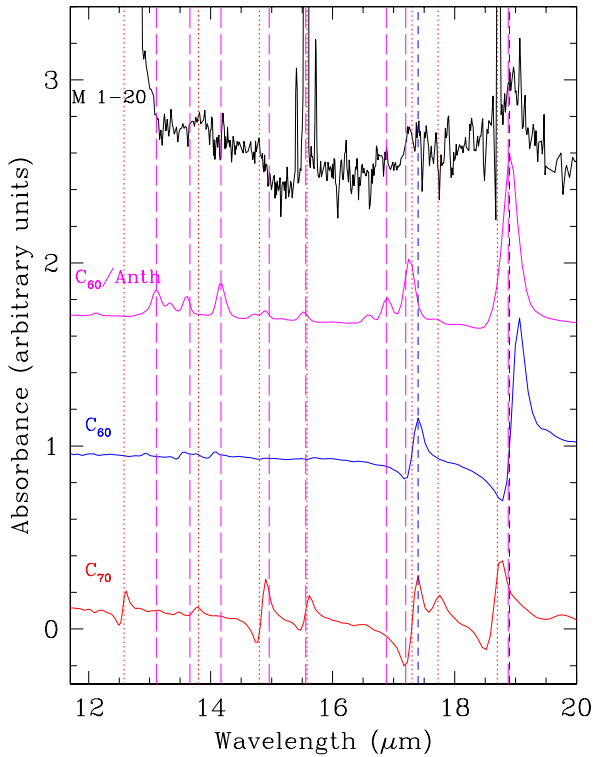}
\caption{Dust continuum subtracted spectra ($\sim$12$-$20 $\mu$m) of PN M
1-20 (Garc\'{\i}a-Hern\'andez et al. 2010) in comparison with the
C$_{60}$/anthracene adduct (magenta), neutral C$_{60}$ (blue), and C$_{70}$ (red)
laboratory spectra (see the legend of Figure 4 for more details). Note that the
C$_{60}$/anthracene adduct laboratory spectrum has been shifted by -0.1 $\mu$m
to show that the 17$\mu$m  feature in M 1-20 seems to be blue-shifted (see
text).} 
\end{figure}

\begin{figure}
\epsfxsize=8truecm
\epsffile{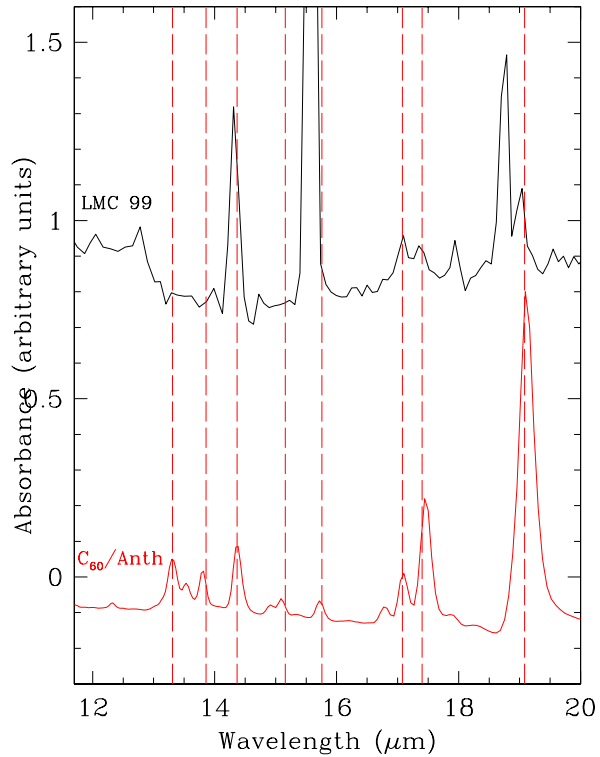}
\caption{Dust continuum subtracted spectra ($\sim$12$-$20 $\mu$m) of the
Magellanic Cloud PN LMC 99 (Garc\'{\i}a-Hern\'andez et al. 2011c) in comparison
with the purified C$_{60}$/anthracene adducts (1:2 initial molar ratio)
spectrum. Note that the C$_{60}$/anthracene adducts laboratory spectrum has been
shifted by $+$0.1 $\mu$m to show that the 17$\mu$m feature in LMC 99 could be
red-shifted (see text).}
\end{figure}

\begin{figure}
\epsfxsize=8truecm
\epsffile{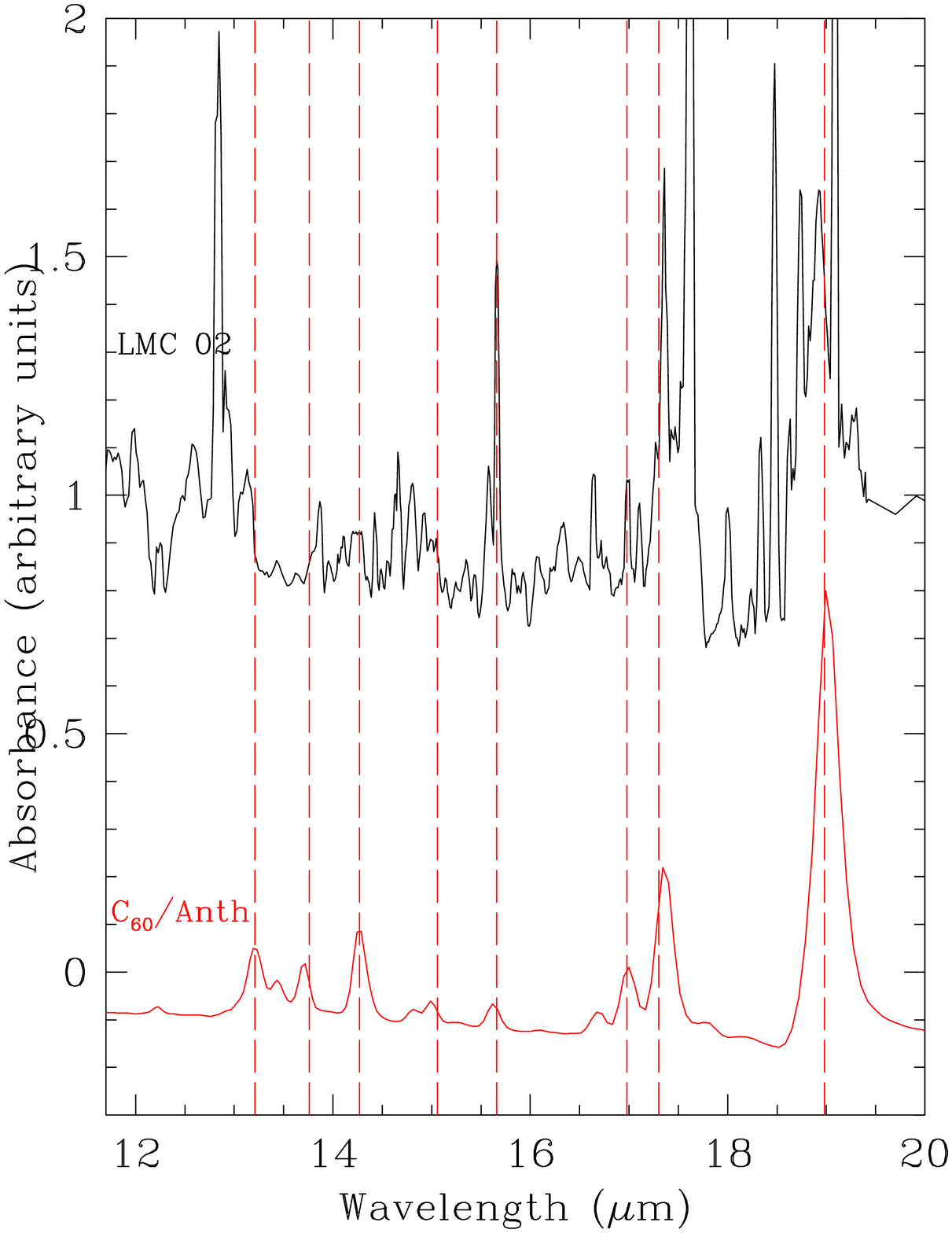}
\caption{Dust continuum subtracted spectra ($\sim$12$-$20 $\mu$m) of the
Magellanic Cloud PN LMC 02 (Garc\'{\i}a-Hern\'andez et al. 2011c) in comparison
with the purified C$_{60}$/anthracene adducts (1:2 initial molar ratio)
spectrum.}
\end{figure}

In summary, the most accepted idea about fullerene formation (at least in PNe)
is that these complex molecular species probably evolved from HAC processing and
present observations of the relative spatial distribution of fullerenes and PAHs
cannot conclusively demonstrate a different location for the latter species.
Thus, fullerenes and PAHs may be mixed in the circumstellar envelopes of PNe,
opening the possibility of forming adducts of fullerenes with PAHs. We have
shown here that fullerenes can at least react with catacondensed PAHs via the
Diels-Alder cyclo-addition reaction. In principle, the smallest PAHs are
expected to be destroyed quite fast in the harsh UV irradiation field present in
PNe or in the ISM. A small number of such fullerene adducts may be formed in
regions where fullerenes and PAHs coexist, although it seems unlikely that they
would be abundant enough to be detected in astrophysical environments. Fullerene
adducts with larger and more stable PAHs (like pericondensed or very large PAHs)
are expected to be more stable towards UV irradiation and hence more abundant in
space. However, the possible Diels-Alder reaction between fullerenes and
pericondensed or very large PAHs has to be explored at laboratory. 

The fullerene C$_{60}$/anthracene adduct reported here is just the simpler
example of the variety of C$_{60}$/PAHs molecules that may be formed. The UV
irradiation of C$_{60}$ and anthracene mixtures - at least in the laboratory
conditions - favors the formation of the C$_{60}$/anthracene mono- and
bis-adducts (Mikami et al. 1998); although other byproducts such as the
anthracene dimer are also formed. In addition, the bis-adduct (with a
decomposition temperature T$_{dec}$=533 K) appears more stable than the
mono-adduct which decomposes at T$_{dec}$=394 K, yielding back free C$_{60}$ and
anthracene (Cataldo et al. 2013). Taking into account that the fullerene
C$_{60}$ temperatures are higher than 400 K for most fullerene PNe
(Garc\'{\i}a-Hern\'andez et al. 2012a), this would imply that
C$_{60}$/anthracene-adducts - if formed and stable/abundant enough in the
UV irradiated environments of PNe - are most likely in the form of trans-1 
bis-adducts.

\section{Comparison with Planetary Nebulae spectra}

As in the case of the C$_{60}$ and C$_{70}$ fullerenes (Iglesias-Groth et al.
2011), we expect the strength and position (and width) of the infrared features
of C$_{60}$/anthracene adducts to be temperature dependent. With this in mind,
we have compared the laboratory FT-IR spectrum of the pure C$_{60}$/anthracene
bis-adducts with the dust continuum subtracted {\it Spitzer Space Telescope}
spectra of the fullerene PNe Tc 1 and M 1-20 (Garc\'{\i}a-Hern\'andez et
al. 2010) in Figures 4 and 5, respectively. M 1-20 displays clear PAH-like
features (e.g., those at 6.2, 7.7, and 11.3 $\mu$m) while Tc 1 show a
fullerene-dominated spectrum with no clear signs of PAH-like features. For
comparison, the laboratory spectra of neutral C$_{60}$ and C$_{70}$ molecules
(Iglesias-Groth et al. 2011) are also shown in Figures 4 and 5. The
laboratory spectra of C$_{60}$/anthracene bis-adducts seem to show {\bf
tentative} resemblance with the Spitzer spectrum of the PN M 1-20. For example,
the substructure of the 13-15 $\mu$m emission complex observed in M 1-20 is
somewhat similar to the C$_{60}$/anthracene features is this spectral region. In
addition, M 1-20 displays a weak feature around 17 $\mu$m that is slightly
blue-shifted (at 16.88 $\mu$m)\footnote{Note that the 17$\mu$m feature in the
hotter PN LMC 99 (see Fig. 6) is red-shifted (at 17.11 $\mu$m). These small
displacements towards the blue and red in M 1-20 and LMC 99, respectively, are
also seen in the other mid-IR features (e.g., at 17.4 and 18.9 $\mu$m), being
consistent with different average temperatures for these sources.} and is weaker
than the 17.3 $\mu$m fullerene feature, something that is also observed in the
C$_{60}$/anthracene bis-adducts spectrum. The 13-15 $\mu$m emission complex and
the weak 17$\mu$m feature are, however, not observed in Tc 1 that also lacks the
PAH-like features.

\begin{figure} 
\epsfxsize=8truecm 
\epsffile{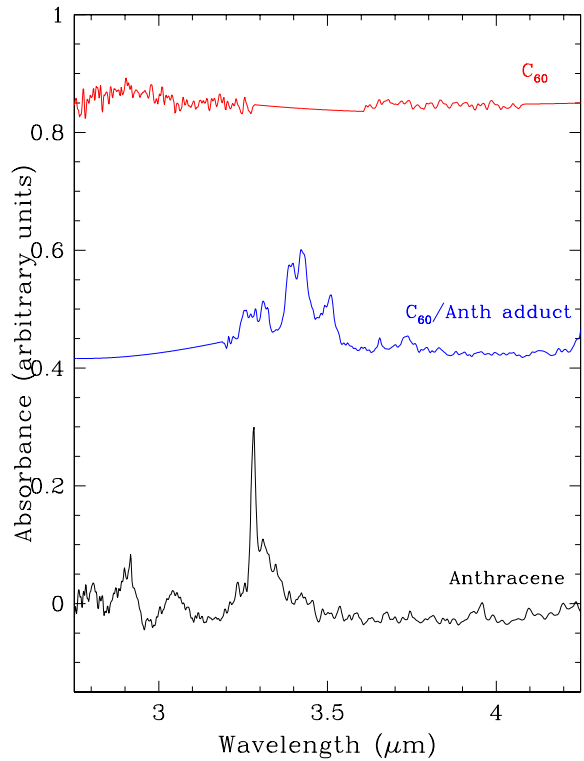}
\caption{Laboratory spectrum ($\sim$3-4 $\mu$m) of fullerene
C$_{60}$/anthracene adducts (in blue) compared with the laboratory spectra of
isolated C$_{60}$ (in red) and anthracene (in black).}
\end{figure}

The still unidentified 17$\mu$m feature is observed in a variety of strong PAH
emitters (e.g., reflection nebulae, H II regions, Herbig Ae/Be stars) that also
show emision features at $\sim$17.4 and 18.9 $\mu$m (e.g., Boersma et al. 2010).
The latter three features are usually accompanied by a much stronger and still
unidentified 16.4 $\mu$m feature. Indeed, by studying the IR spectra of HAC
nano-particles samples, Duley \& Hu (2012) suggest that the set of four IR
emission lines previously identified with C$_{60}$ in objects that also show the
16.4 $\mu$m feature and other strong PAH bands arise from proto-fullerenes
rather than C$_{60}$ (see also Garc\'{\i}a-Hern\'andez et al. 2012b). Our IR
laboratory spectra of C$_{60}$/anthracene adducts, show, however, that these
four features at $\sim$7.0, 8.4, 17.3, and 18.9 $\mu$m (and other IR features
coincident with C$_{70}$ transitions) may also arise from fullerene-adducts. It
should be noted that the 16.4 $\mu$m feature is not present in fullerene PNe and
the PAH bands are very weak (Garc\'{\i}a-Hern\'andez et al. 2012a). In addition,
some fullerene PNe still displays the 17$\mu$m feature and/or a rich IR spectrum
with several unidentified features (Garc\'{\i}a-Hern\'andez et al. 2011c). In
Figures 6 and 7, we attempt a tentative identification of these unidentified
features by comparing our IR laboratory spectra of pure C$_{60}$/anthracene
bis-adducts with the {\it Spitzer Space Telescope} observations of PNe LMC 99
and LMC 02. LMC 99 is the fullerene PNe with the strongest PAH bands while LMC
02 displays an IR spectra richer than other fullerene PNe. It can be seen that
LMC 99 displays the $\sim$17.0, 17.3, and 18.9 $\mu$m features together with the
lack of the 16.4 $\mu$m feature (Fig. 6). Also, some C$_{60}$/anthracene adduct
laboratory features (e.g., at 13.2, 13.8, 14.3, and 17.0 $\mu$m) find a
counterpart in the LMC02's spectrum (Fig. 7). However, other IR features (e.g.,
at 16.6, 18.0, and 18.4 $\mu$m) that are observed in LMC 02, are not present in
the laboratory spectra. In any case, this comparison suggests that other
fullerene-related molecules such as bigger fullerene-adducts could be potential
candidate carriers for these peculiar and still unidentified IR features seen in
some fullerene PNe.

Our comparison of the fullerene-adducts laboratory spectra with the PNe spectra
do not permit us to unambiguously infer the presence of these fullerene-based
molecules in PNe. As we have mentioned above, we expect the exact wavelength
positions (also the relative intensities and widths) of the C$_{60}$/anthracene
adduct spectral features to be temperature dependent (e.g., $\pm$0.1 $\mu$m in
wavelength; e.g., Iglesias-Groth et al. 2011, 2012). A detailed study on the
dependence of the IR bands of C$_{60}$/anthracene adducts over a wide range of
temperatures is planned, which will be useful for the search and identification
of fullerene-adducts in space. However, it is clear that fullerene-adducts
display mid-IR features strikingly similar to the features that have been
previously attributed to C$_{60}$ (and C$_{70}$). Our work show that
C$_{60}$/anthracene adducts may be indistinguishable from C$_{60}$ (and
C$_{70}$) on the basis of the {\it Spitzer} mid-IR ($\lambda$ $>$ 5 $\mu$m)
spectra alone. A possible spectroscopic test to discriminate between
fullerene-adducts and isolated C$_{60}$ (and C$_{70}$) molecules could be
offered by the 3-4 $\mu$m spectral region. The laboratory spectra show that the
3-4 $\mu$m region may be the best spectral range to distinguish between
fullerene-adducts and neutral C$_{60}$ (and C$_{70}$). This is shown in Figure 8
where we compare the 3-4 $\mu$m laboratory spectrum of C$_{60}$/anthracene
adducts with those of isolated C$_{60}$ and anthracene. C$_{60}$/anthracene
adducts display the typical aromatic C-H bands around 3.3 $\mu$m and a
series of aliphatic C-H bands at 3.39, 3.43, and 3.52 $\mu$m, which are not
present in the C$_{60}$ spectrum. These aliphatic C-H bands appear cleanly
separated at the resolution (R$\sim$600) of our experimental set-up. In
principle, the possible carrier (e.g., C$_{60}$ vs. C$_{60}$/PAHs adducts) of
the mid-IR features seen in fullerene PNe by {\it Spitzer} could be
elucidated through high-resolution (R$>$500) spectroscopic observations in the
3-4 $\mu$m range. However, the appearance in our laboratory
C$_{60}$/anthracene adducts spectrum of C-H stretching bands that are typical
for CH$_{2}$ and CH$_{3}$ groups (see below) is not yet completely understood.
The aliphatic C-H bands in the 3.4-3.5 $\mu$m spectral region (e.g.,
asymmetric and symmetric stretching vibrations of aliphatic CH$_{2}$ and
CH$_{3}$ groups) are also very typical of several carbonaceous materials (see
e.g. Pendleton \& Allamandola 2002). For example, Pendleton \& Allamandola
(2002) show that even the measured positions of the 3.4-3.5 $\mu$m bands due to
aliphatic CH$_{2}$ and CH$_{3}$ groups in HAC are quite close to those in
fullerene-adducts. Therefore, a careful analysis of the whole infrared spectra
as well as a much better understanding of our C$_{60}$/anthracene adducts
laboratory spectrum (e.g., the nature of the 3-4 $\mu$m C-H stretching bands)
would be needed, in order to fully disentangle the possible carrier of these
emission bands. 

\section{Concluding remarks}

The detection of fullerenes in evolved stars as common as our Sun has opened the
exciting possibility that other forms of carbon such as hydrogenated fullerenes,
fullerene-adducts, and multishell fullerenes (buckyonions) may be widespread in
the Universe. In particular, if fullerenes and PAHs evolve from the
decomposition of HACs, then both species may be mixed in the circumstellar
envelopes of fullerene PNe, forming fullerene/PAHs adducts (at least with
catacondensed PAHs). Our IR laboratory spectra of C$_{60}$/anthracene adducts
show that C$_{60}$ (and C$_{70}$) and other unidentified infrared emission
features could also arise from fullerene-adducts. 

Interestingly, Bernard-Salas et al. (2012) could not explain their observations
of C$_{60}$ fullerenes (spatial distribution, relative strenght of the features)
in several PNe and they suggest that other emitting species (e.g., fullerene
clusters or nanocrystals) may be present. In this paper, we present some
evidence that fullerene-based molecules such as fullerene/PAH adducts could be
present in PNe and may contribute to the mid-IR features attributed to C$_{60}$.
This could be a more natural explanation because fullerenes and PAHs may coexist
in PNe. The possible presence of other fullerene-based molecules (e.g.,
fullerenes bigger than C$_{60}$ and buckyonions) in fullerene PNe is also
suggested by the special characteristics of the DIBs seen towards Tc 1 and M
1-20 (Garc\'{\i}a-Hern\'andez \& D\'{\i}az-Luis 2013). 

We speculate that fullerene/PAHs adducts - if formed under astrophysical
conditions and abundant enough - could contribute to the infrared emission
features seen in fullerene-containing environments. The C$_{60}$/anthracene
adduct presented in this paper is just a first example of a family of
C$_{60}$/PAHs molecules, which could be present in various space environments.
In this context, fullerene adducts with larger catacondensed PAHs (e.g.,
C$_{60}$/pentacene) or even adducts with pericondensed or very large PAHs
could represent potential candidate carriers of the still unidentified infrared
features seen in fullerene PNe but a definitive answer requires further
laboratory and observational efforts.

\section*{Acknowledgments}

The present research work has been supported by grant AYA2007$-$64748 Expte.
NG$-$14$-$10 of the Spanish Ministerio de Ciencia e Innovacion. D.A.G.H. and
A.M. also acknowledge support provided by the Spanish Ministry of Economy and
Competitiveness under grant AYA$-$2011$-$27754.


\begin{thebibliography}{99}
\bibitem[\protect\citeauthoryear{}{}]{b} Allamandola, L. J., Sandford, S. A., Tielens, A. G. G. M., Herbst, T. M. 1992, ApJ, 399, 134
\bibitem[\protect\citeauthoryear{}{}]{b} Bernard-Salas, J. et al. 2012, ApJ, 757, 41 
\bibitem[\protect\citeauthoryear{}{}]{b} Bettens, R. P. A., \& Herbst, E. 1996, ApJ,468, 686
\bibitem[\protect\citeauthoryear{}{}]{b} Boersma, C. et al. 2010, A\&A, 511, 32
\bibitem[\protect\citeauthoryear{}{}]{b} Briggs, J. B., \& Miller, G. P. 2006, Comptes Rendus Chimie 9, 916
\bibitem[\protect\citeauthoryear{}{}]{b} Cami, J., Bernard-Salas, J., Peeters, E., \& Malek, S. E. 2010, Science, 329, 1180
\bibitem[\protect\citeauthoryear{}{}]{b} Cami, J., Bernard-Salas, J., Peeters, E., \& Malek, S. E. 2011, in The Molecular Universe, IAU Symposium 280, p. 216
\bibitem[\protect\citeauthoryear{}{}]{b} Cataldo, F., Garc\'{\i}a-Hern\'andez, D. A., \& Manchado, A. 2013, Fullerenes Nanot. Carbon Nanostruct. (in press) 
\bibitem[\protect\citeauthoryear{}{}]{b} De Vries, M. S. et al. 1993, Geochim. Cosmochim. Acta, 57, 933
\bibitem[\protect\citeauthoryear{}{}]{b} Duley, W. W., \& Hu, A. 2012, ApJ, 745, L11
\bibitem[\protect\citeauthoryear{}{}]{b} Garc\'{\i}a-Hern\'andez, D. A. et al. 2010, ApJ, 724, L39 
\bibitem[\protect\citeauthoryear{}{}]{b} Garc\'{\i}a-Hern\'andez, D. A. et al. 2011a, ApJ, 729, 126
\bibitem[\protect\citeauthoryear{}{}]{b} Garc\'{\i}a-Hern\'andez, D. A. et al. 2011b, ApJ, 739, 37
\bibitem[\protect\citeauthoryear{}{}]{b} Garc\'{\i}a-Hern\'andez, D. A. et al. 2011c, ApJ, 737, L30
\bibitem[\protect\citeauthoryear{}{}]{b} Garc\'{\i}a-Hern\'andez, D. A. et al. 2012a, ApJ, 760, 107
\bibitem[\protect\citeauthoryear{}{}]{b} Garc\'{\i}a-Hern\'andez, D. A. et al. 2012b, ApJ, 759, L21
\bibitem[\protect\citeauthoryear{}{}]{b} Garc\'{\i}a-Hern\'andez, D. A. \& D\'{\i}az-Luis, J. J. 2013, A\&A, 550, L6
\bibitem[\protect\citeauthoryear{}{}]{b} Gielen, C., Cami, J., Bouwman, J., Peeters, E., \& Min, M. 2011, A\&A, 536, 54
\bibitem[\protect\citeauthoryear{}{}]{b} Hirsch, A., \& M. Brettreich, M. 2005, Wiley-VCH, Weinhem
\bibitem[\protect\citeauthoryear{}{}]{b} Iglesias-Groth, S., Cataldo, F., \& Manchado, A. 2011, MNRAS, 413, 213 
\bibitem[\protect\citeauthoryear{}{}]{b} Iglesias-Groth, S., Garc\'{\i}a-Hern\'andez, D. A., Cataldo, F., Manchado, A. 2012, MNRAS, 423, 2868
\bibitem[\protect\citeauthoryear{}{}]{b} Komatsu, K. et al. 1993a, Fullerenes Nanot. Carbon Nanostruct., 1, 231
\bibitem[\protect\citeauthoryear{}{}]{b} Komatsu, K. et al. 1993b, Tetrahedron Lett., 34, 8473
\bibitem[\protect\citeauthoryear{}{}]{b} Komatsu, K. et al. 1999, Fullerenes Nanot. Carbon Nanostruct, 7, 609
\bibitem[\protect\citeauthoryear{}{}]{b} Kroto, H. W. et al. 1985, Nature, 318, 162
\bibitem[\protect\citeauthoryear{}{}]{b} Micelotta, E. R., Jones, A. P., Cami, J. et al. 2012, ApJ, 761, 35
\bibitem[\protect\citeauthoryear{}{}]{b} Mikami, K. et al. 1998, Tetrahedron Lett., 39, 3733
\bibitem[\protect\citeauthoryear{}{}]{b} Leger, A., \& Puget, J. L. 1984, A\&A, 137, L5
\bibitem[\protect\citeauthoryear{}{}]{b} Peeters, E., Tielens, A. G. G. M., Allamandola, L. J., Wolfire, M. G. 2012, ApJ, 747, 44
\bibitem[\protect\citeauthoryear{}{}]{b} Pendleton, Y. J., \& Allamandola, L. J. 2002, ApJS, 138, 75
\bibitem[\protect\citeauthoryear{}{}]{b} Roberts, K. R. G., Smith, K. T., \& Sarre, P. J. 2012, MNRAS, 421, 3277
\bibitem[\protect\citeauthoryear{}{}]{b} Rubin, R. H., Simpson, J. P., O'Dell, C. R., et al. 2011, MNRAS, 410, 1320
\bibitem[\protect\citeauthoryear{}{}]{b} Scott, A. D., Duley, W. W., \& Pinho G. P. 1997, ApJ, 489, L193
\bibitem[\protect\citeauthoryear{}{}]{b} Sellgren, K., Werner, M. W., Ingalls, J. G. et al. 2010, ApJ, 722, L54
\bibitem[\protect\citeauthoryear{}{}]{b} Wang, G. W., Chen, Z. X, Murata, Y., \& K. Komatsu, K. 2005, Tetrahedron, 61, 4851
\bibitem[\protect\citeauthoryear{}{}]{b} Yurovskaya, M. A., \& Trushkov, I. V. 2002, Russian Chem. Bull. Int. Ed. 51, 367
\bibitem[\protect\citeauthoryear{}{}]{b} Zhang, Y. \& Kwok, S. 2011, ApJ, 730, 126
\end{thebibliography}
\end{document}